# ReaxFF Simulations of Self-Assembled Monolayers on Silver Surfaces and Nanocrystals

A. Lahouari,[1] J.−P. Piquemal,[1] and J. Richardi[1]*

*1 Sorbonne Université, Laboratoire de Chimie Théorique, UMR 7616 CNRS , Paris 75005 , France*

*Email johannes.richardi@sorbonne−universite.fr

**Abstract**

The self−assembled monolayers of alkanethiolates on Ag (111) surfaces and nanoparticles are studied using molecular dynamics. Reactive force fields allow simulations of very large systems, such as nanoparticles of 10 nm. Stable $(\sqrt{7} \times \sqrt{7})R19.1°$ assemblies are obtained as experimentally observed for these systems. Only nanoparticles smaller than 4 nm show a spontaneous restructuring of the metallic core. The preferred adsorption site is found to be in an on−top position, which is in good agreement with recent X−ray absorption near edge structure experiments. Moreover, similar distances between the sulfur headgroups are found on the facets and edges.

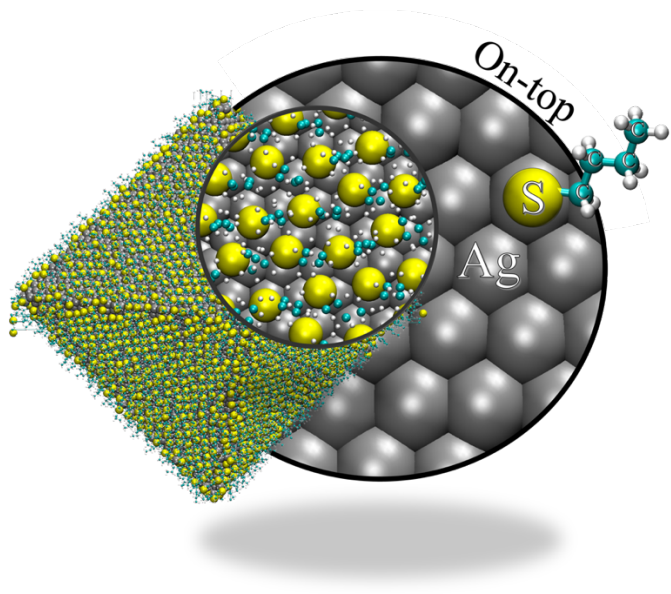

# 1. Introduction

Silver nanoparticles (NPs) are characterized by unique optical, electrical, thermal, and biological properties. 1,2 In particular, they are employed as antibacterial agents in tissues, food products, and health care.3 Thus, silver nanoparticles have exceptional antimicrobial activity. Moreover, their surface plasmon resonance gives them enhanced absorption and scattering of light. Therefore, nanosilver is nowadays one of the most widely used nanomaterials. For many applications, silver NPs are protected by ligands, such as thiolates or amines. These ligands form self−assembled monolayers (SAMs) on the NP surface. SAMs are usually used to avoid nanoparticle's coalescence. But they are also useful to modify the properties of NPs, such as their solvability in specific solvents. Alkanethiolates are, for example, used to tune the antibacterial properties of silver NPs.3 Their presence modulates the $Ag^+$ ion release and helps to avoid nondesirable side effects. Recently, the synthesis of atomically defined thiolate−protected nanoclusters was achieved, opening up many possible applications. 4,5

Our study focuses on the alkanethiolates SAM, which form on Ag (111) surfaces and NPs. Experimentally, a $(\sqrt{7} \times \sqrt{7})R19.1°$ SAM (denoted as 77 SAM in the following) is observed. 6,7 This geometry is denoted by Wood's notation.8 It means that the thiolate head groups form a hexagonal assembly that is extended by $\sqrt{7}$ and turned by 19.1° with respect to the layer of Ag atoms on the Ag (111) surface. Moreover, experimental and theoretical studies have also shown that, due to the presence of thiolates, the silver surface can be restructured. 6,7 In this case, a silver atom appears between three sulfur atoms of the thiolates due to bonds breaking on the surface. For Au (111) surfaces, two superstructures are found, c (2 × 4) and $(\sqrt{3} \times \sqrt{3})R30°$ SAMs (denoted as 33). The former is usually caused by the formation
of staples, which corresponds to a restructuration of the surface and is found in the majority. 9,10 For this first study, however, we focus on nonrestructured silver facets.

Silver NPs protected with thiolates have been widely studied by experiments in the last years 11−17 using techniques like EXAFS (extended X−ray absorption fine structure), XANES (X−ray absorption near edge structure), or XPS (synchrotron−based X−ray photoelectron spectroscopy). These experiments led to two central questions. First, in some experiments, the presence of silver sulfide was observed between the Ag core and the surrounding thiolates. 11,12,17 However, other experiments found only a small amount of silver sulfide. 13 − 17 The eventual origin of silver sulfides has not been elucidated until now.17 Second, very recent experiments indicate an on−top adsorption site for the thiolates. This is in contradiction to recent theoretical density functional theory (DFT) studies, which predict a preference for the bridge position on flat silver surfaces and NPs. 18,19

The aim of this paper is to apply the recently developed ReaxFF20 force field (FF) in order to perform molecular dynamics simulations of thiolate SAMs on silver. Recently, several aspects of silver NPs have been studied using FF simulations 21−25 and DFT calculations.20 The FF studies handle aspects that are not in close relation with thiolate−coated NPs, as investigated here. These aspects include the interaction between silver NPs and proteins,22 the interaction with amines,23 and the formation of hollow silver NPs under laser irradiation.24 The DFT study investigates the adsorption of methanethiolates on both silver nanoparticles and (111) surfaces. The results revealed that bridge sites are the most thermodynamically stable adsorption locations, emphasizing their significance. Additionally, it was observed that smaller nanoparticles display increased reactivity, accompanied by significant adsorption energies, further highlighting their role as reactive platforms.

However, to the best of our knowledge, there is only one study that focused on thiolates Ag (111) surfaces.25 These simulations do not succeed in reproducing the 77 SAM on silver that was experimentally found. Moreover, no simulation of silver NPs with thiolates has been published. The use of ReaxFF may improve this situation for three reasons.

First, classical force fields using, for example, Lennard−Jones parameters naturally lead to hollow sites as preferred adsorption sites (see, for example, the study in ref 26). This is because the hollow sites enable more contact between the adsorbing atom and the metallic atoms. Therefore, such FFs have difficulties simulating 77 SAMs characterized by the adsorption of bridges and on−top sites. Within the ReaxFF, the Ag–S interaction is handled as a bond, and angle−dependent interactions enable stabilization of the adsorption of bridge and on−top sites compared to the hollow sites. Thus, we will show here that the new FF is able to reproduce the 77 SAM on Ag (111) planes in agreement with the experiment.

Second, the silver surface may be restructured. This can only be simulated with reactive FFs, allowing for the breaking of the Ag–Ag bonds; this would be the case for ReaxFF. Moreover, ReaxFF enables us to study if the thiolate is decomposed during adsorption, which may explain the formation of silver sulfide.

Third, with the help of massively parallel computing, ReaxFF can investigate NPs made of a hundred thousand atoms in contrast to DFT methods. Thus, the largest system studied here is the butanethiolate−coated NP of 10 nm, made of 57 826 atoms. Moreover, the evolution with temperature can also be studied, which implies simulation lengths up to 4 ns. Such ReaxFF has already been applied for several metallic NPs made of cobalt, copper, and gold. 27− 31 In particular, for gold NPs; it predicts staple formation and on−top adsorption sites in full agreement with experiments. 31

Here, the ReaxFF is used to calculate the properties of silver surfaces and NPs (with diameters ranging from 2 to 10 nm) coated with thiolates. To study the influence of chain length, methanethiolates and butanethiolates are used. While methanethiolates have been widely studied on flat silver surfaces, for NPs, usually longer chains with at least butane have been used. For the sake of comparison, we nevertheless investigated methanethiolate on NPs. In this first study, we start from nonrestructured surfaces to see if the restructuring spontaneously happens, as previously observed for gold NPs with thiolates. We also ignored the presence of the solvent. It is expected that the solvent plays a minor role in the adsorption properties, which are mainly governed by the strong Ag–S interactions. However, it might be nonnegligible for the energy difference between 77 and 33 SAMs, which should be investigated in the future.

The study focuses on nanocrystals of octahedral shape, which have been chosen since their facets are made of Ag (111) planes. This enables a direct comparison with the results obtained for the Ag (111) surfaces. Moreover, NPs of octahedral shape are often observed in experiments. 32

The structure of this article is as follows: First, we present the ReaxFF approach and the simulation protocol. We will then introduce the new analysis tools, SAMmaker and SAMfinder, that were specifically developed for this study. They allow the construction and finding of SAMs on metallic surfaces and NPs. Then, the stability of the thiolate SAMs on the Ag(111) surfaces is investigated. Finally, we turn to the predicted properties of the SAMs on NPs.

## 2. Methods

2.1. The ReaxFF model. The reactive force field method proposed by van Duin et al.33 enables the formation and breaking of bonds during classical molecular dynamics simulations. The formalism is described in detail in the paper by Chenoweth et al.34 Here, we use the recently developed FF for AgSCH.20 It is able to reproduce the DFT values for the average distances between silver atoms in NPs with a precision of 0.05 Å. It also correctly describes the energetical and geometric properties of thiolates on an $Ag_{20}$ pyramid arrangement.20 The charges of the atoms are obtained through the usual electronegativity equalization method. 35,36

2.2. Simulation Protocol. Molecular dynamics simulations are carried out in the NVT ensemble using a timestep of 0.25 fs. The temperature is controlled using the Berendsen thermostat with a damping constant of 5 fs. The reliability of the Berendsen thermostat has been verified through simulations using the NVT Nose–Hoover thermostat, yielding identical results within the statistical margin. The parallel program LAMMPS is used for all simulations. 37,38 First, an equilibration is performed by increasing the temperature from 0 to 300 K. The simulation is then continued at 300 K to obtain better converged properties. To test correct equilibration and convergence, three runs are performed:

- Equilibration: 0.3 ns, convergence at 300 K: 0.3 ns

- Equilibration: 1 ns, convergence at 300 K: 1 ns

- Equilibration: 3 ns, convergence at 300 K: 1 ns

To obtain standard errors for all properties, three independent simulations are carried out, starting from different initial atomic positions for the SAM position with a shift of 0.05 Å. It has been found that in the case of NPs, an alkane chain strictly perpendicular to the silver surface has to be avoided since this leads to a strong reduction of the SAM frequency of about 30%. Indeed, the alkane chains go together and move the S atoms from their initial positions corresponding to the chosen SAM.

2.3. Set up Tools. A new Python tool called SAMmaker was developed to set up SAMs on flat surfaces and NPs. It has been included in the Python library NATOMOS. How this tool works is explained in the Supporting Information.

We observed that it is very important to have a compact initial layer of alkane without strong repulsion to obtain stable SAMs. Therefore, all ligands are slightly declined and oriented in the same directions. Moreover, orientations with very small repulsions were searched for. Figure 1 shows typical snapshots obtained after equilibration for 77 and 33 SAM of thiolates on the Ag(111) surface.

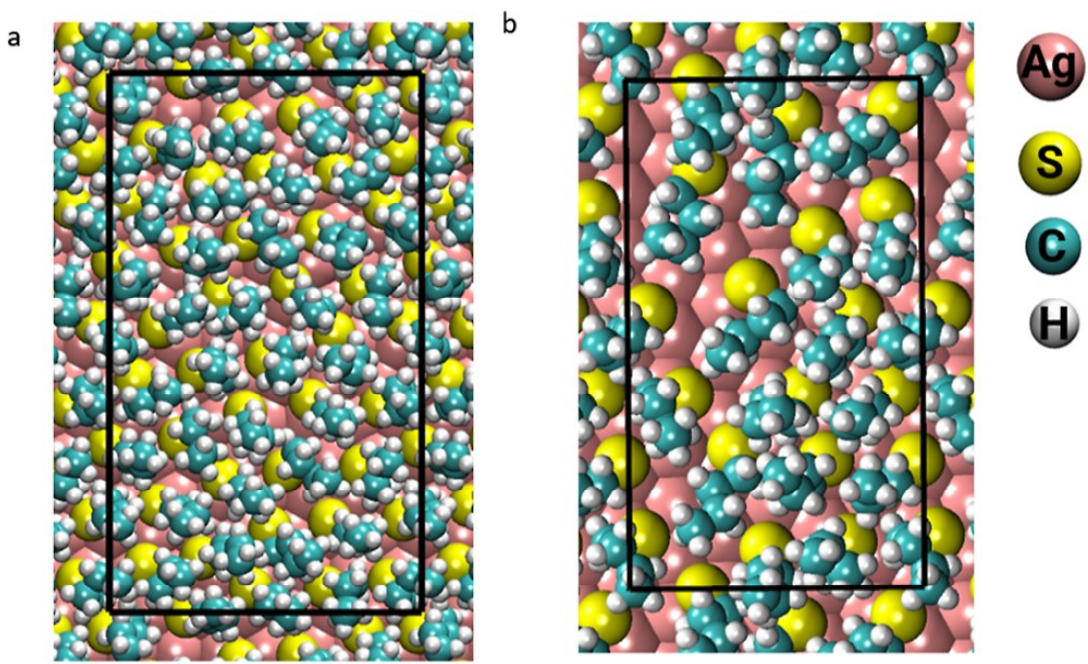

Figure 1: Snapshots of a SAM of on Ag (111) at 300 K after an equilibration of 0.3 ns. (a) 77 SAM, and (b) 33 SAM. The simulation box for each system is represented in black.

To set up the NPs, the SAMs obtained with SAMmaker are transferred to the NP facets using the NATOMOS tool NTM_setup, which has been described in detail in a previous paper.31 It is interesting to note that there are differences in the stability of SAMs for NPs compared to surfaces. Thus, starting from an orientation of the methanethiolate with the S–C bond perpendicular to the plane led to a well−defined SAM on the surface. However, when the same orientation is used for the NPs, the SAM becomes unstable during the simulation. This shows the importance of well−defining the initial orientation to obtain a large degree of SAM.

2.4. Analysis Tools. The SAMs were analyzed with the help of the Python tool NTM_ana. It was described in detail in a previous paper.31 It determines the following properties studied in this paper:

- the number of neighbors for the silver atoms,
- the locations and types of the adsorption sites occupied by the ligands,
- the average distances of neighboring ligand head groups.

Please note that two atoms are counted as neighbors whenever the distance is smaller than the position of the minimum after the first peak of the pair distribution function (minimum for Ag–Ag: 3.75 , Ag–S: 2.92, and S–S: 5.5 Å).

Here, a new function, SAMfinder, has been included in Ntm_ana. It allows the determination of the percentage of ligands participating in a SAM characterized as 33 or 77, for example. For this purpose, a percolation method is used, which is explained in detail in the Supporting Information. All properties obtained by the analysis are shown in Tables S1 to S4 in the.

## 3. Results and Discussion

3.1. Thiolate SAMs on Ag (111) Surfaces. Simulations on two different thiolate SAMs on an Ag (111) plane were carried out (see Figure 1 ). On one hand, the 77 SAM experimentally observed on nonrestructured Ag (111) surfaces is studied. On the other hand, we also investigated the 33 SAM for silver, usually observed on an Au (111) when no staple formation occurs. The second SAM was chosen for the following reasons: The Ag (111) and Au (111) surfaces have the same geometry, leading to the same distances between the ligands in a given SAM. Moreover, staple formation, such as for Au (111), is not observed on Ag (111). Therefore, the 33 SAM without staples should also be possible for steric reasons on Ag (111). A successful force field should, however, predict that the 77 SAM is more stable than the 33 SAM, in agreement with experiments.

We now study the simulations starting from 77 and 33 SAMs as a function of the temperature. Figure 1 shows snapshots of both SAMs obtained at 300 K. First, the stability of the SAMs during the simulation is studied. Then, their energies and entropies are compared, and finally, their properties, such as the occupation of adsorption sites, are discussed. The temperature is continuously increasing. Every 50 K, the simulation is stopped and continued at a fixed temperature for 30k, 100k, and 300k steps corresponding to the three equilibrations (0.3, 1, and 3 ns). This is done to ensure that the results are well−converged. Only the last third of these runs at fixed temperatures are used to calculate the SAM frequencies and energies.

3.1.1. Stability of the SAMs on Silver Surfaces. Figure 2 plots the percentage of thiolates belonging to 77 and 33 SAM for methane and butanethiolate obtained for the longest simulation run (3 ns), as determined with the module SAMfinder. The results for both 77 and 33 SAMs at different simulation times are shown in the Supporting Information (Figure S1). Please note that parts a and b show the frequencies obtained from two different series of simulations starting from 77 and 33 SAM, respectively.

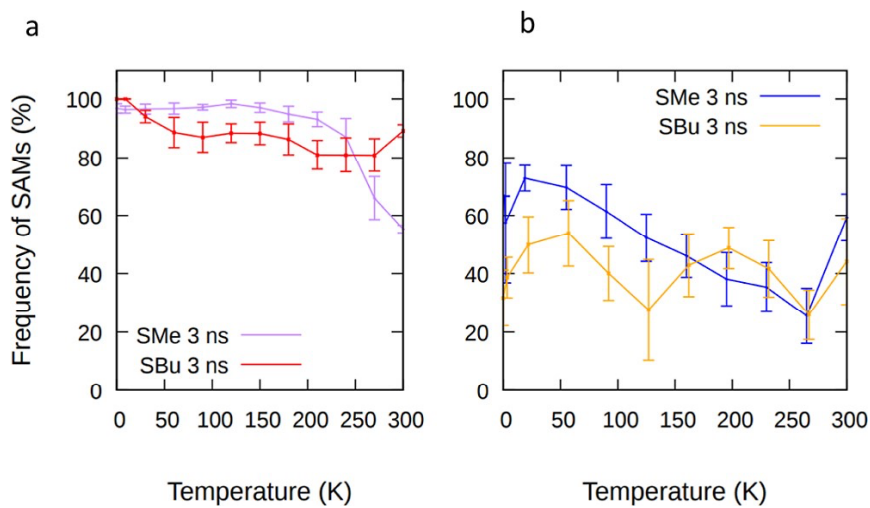

Figure 2: Frequencies of 77 (a) and 33 (b) SAMs of methane and butanethiolates on the Ag(111) surface for an equilibration step of 3 ns corresponding to an increase in the temperature.

In the case of butanethiolate, the 77 SAM occupied nearly the total surface up to 300 K, independent of the simulation time. This is also found for methanethiolate until 250 K. However, at higher temperatures, the percentage of methanethiolates in 77 SAMs decreases. This becomes more pronounced when the simulation time used for the equilibrations increases (see Figure S1). To investigate the stability of the 77 SAM at 300 K, the simulation run is continued for 1 ns at that temperature (see Figure S2). We observed a complete disappearance of the 77 SAM for methanethiolate in contrast to butanethiolate, which rests stable. Studying the snapshots at 300 K (see Figure S3), we found a desorption of the methanethiolates, which is not observed in experiments. Thus, it shows the limit of our force field. Since we want to study stable SAMs, we only considered simulations before this happens (the first 10 000 steps at 300 K). Figure 2 also shows the frequency of 33 SAMs starting from a perfect 33 SAM, where the frequency of SAM rapidly decreases to about 20% for the methane and stays stable for the butane with some fluctuation during the simulation. Note that the results are the same within the statistical error for the 1 and 3 ns simulations. This indicates that the results are well converged. The only exception is the simulation for methanethiolate at temperatures larger than 200 K, which has been explained by a very slow desorption of the thiolates. To be sure that a similar process does not happen for butanethiolate, we have continued the simulation at any temperature for 3 ns. The results are stable within the statistical errors.

3.1.2. Energy and Entropy of the SAMs. In order to compare the stability of both SAMs, we first calculate their binding energies per thiolates, defined as:

$$\Delta E_{bind} = \frac{1}{n_{Thio}}(E^{M-SAM} - E^{Metal} - n_{Thio}E^{Thiolate})$$

where $E^{M-SAM}$ corresponds to the energy of the system, which corresponds to the SAMs in contact with the metal layer. $E^{METAL}$ and $E^{THIOLATE}$ are, respectively, the energies of the metal layer and of the single thiolate molecule. $n_{THIO}$ corresponds to the number of thiolates present on the surface. We show in the following the differences in binding energy between 77 and 33 SAMs. By calculating this difference, one can easily see that the calculation of $E^{THIOLATE}$ is not necessary, which allows us to decrease the possible errors that can emerge with the simulation of a small system using ReaxFF. Please note that the binding energy difference is negative, implies that the 77 SAM is energetically preferred.

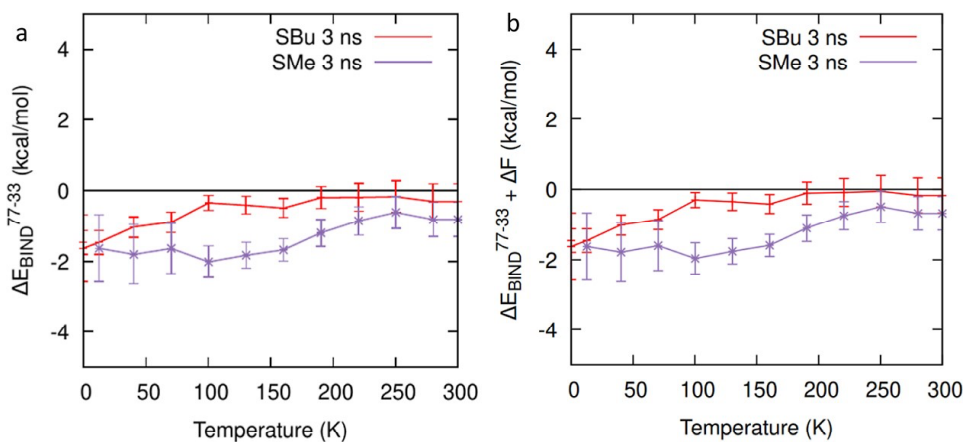

Figure 3: a) Is the difference of binding energies between the 77 and the 33 SAMs and (b) the entropy is added, as a function of the temperature for both butane and methanethiolates on the Ag (111) surface. Simulation length: 3 ns.

Figure 3 displays the evolution of the binding energies as a function of the temperature for methane and butanethiolates on silver for 3 ns. In Figure S4, the results for 0.3 and 1 ns are shown. It is obvious that sufficiently long simulation runs must be carried out to obtain stable results.

In the case of methanethiolate, Figure 3 shows that the binding energy difference varies between −2 and −1 kcal mol$^{-1}$. This indicates that 77 SAM is energetically favored. However, for butanethiolate, the binding energy difference is between −2 and 0 kcal mol$^{-1}$. Hence, the energetic stabilization of the 77 conformations is not proved in this case.

In order to determine which SAM is the most stable, we need to consider the influence of entropy. It can be estimated from statistical mechanics using the partition function of translation. The ligand molecule can move within the space left by the neighboring ligands. For the 33 SAM, this space is larger due to a larger surface per thiolate than that for the 77 SAM. It is, therefore, expected that the 33 SAM is entropically favored. In Figure 3 b, the estimation of the entropy difference is added to show how it may influence the stability of the SAMs. The evaluation of the free energy part due to entropy at 300 K gave 0.149 kcal mol$^{-1}$ between the 77 and the 33 SAM. This shows that the entropy favors 33 SAMs, as expected. However, in the case of methanethiolate, the binding energy is much larger than the estimated free energy due to entropy. This means that the 77 SAM is the more stable assembly for our force field, which is in good agreement with the experiment. As discussed above, this conclusion cannot be made in the case of butanethiolate.

3.1.3. The Properties of the 77 SAMs. We now look at the properties of the 77 SAMs, which will be important for the comparison with the results obtained in the following section for the NPs.

It is important to note that during all simulations, no spontaneous decomposition of the ligands is observed. We first study the number of Ag neighbors around the silver atom. The expected numbers of 12 and 9 neighbors for the bulk and surface atoms, respectively, were obtained as shown in Figure S5. No spontaneous restructuring of the metallic layer is observed. DFT theory and experiments, however, have shown such a restructuration. [6,7] This may mean that such a restructuring is too slow to take place in our simulations. In future simulations starting from a restructured surface, it is planned to study their stability and properties in comparison with the nonrestructured layer investigated here.

We now turn to the occupation frequencies of adsorption sites shown in Figure S6 for butanethiolates. A comparison of the results obtained after 1 and 3 ns shows agreement within the statistical accuracy, which indicates that the results are well converged. A similar occupation of the on−top (52%) and bridge (48%) sites is found where the S atom is in contact with one or two silver atoms, respectively. These sites are typically occupied in 77 SAMs when studied by DFT.[6] Please note that we found that these sites are usually occupied in a way so that the adsorbing atom is shifted from the ideal on−top or bridge site.

Finally, the average distance between the sulfur head groups of the ligands was determined (Figure S7). A value of 4.4 Å is found, which corresponds to the one expected for the 77 SAM.

3.2. Thiolate SAMs on Silver Nanoparticles. Figure 4 shows the typical snapshot obtained for simulations of 2 and 10 nm NP with butanethiolate at 300 K (1 ns). Figure S8 presents snapshots of 4 nm. The NPs of 2 and 4 nm are also shown without the ligands to make any restructuration visible. No spontaneous decomposition of the thiolates was observed. Therefore, the formation of silver sulfide observed in some experiments 11,12,17 cannot be explained by these experiments. However, we cannot exclude that a restructured silver layer may lead to this kind of decomposition, which should be studied in the future. Also, other ligands such as allyl thiolate (for which silver sulfide has been often observed 12,17 ) should be tested. We will now first study whether the 77 SAM is also stable on NPs. In the following, the properties, such as the occupation of adsorption sites and the distance between the sulfur atoms, are discussed for the NPs. The sulfur atoms on the edges in Figure 4 may give the impression of being separated from the alkane chains. This is not the case. They are oriented only to the facets of the NPs.

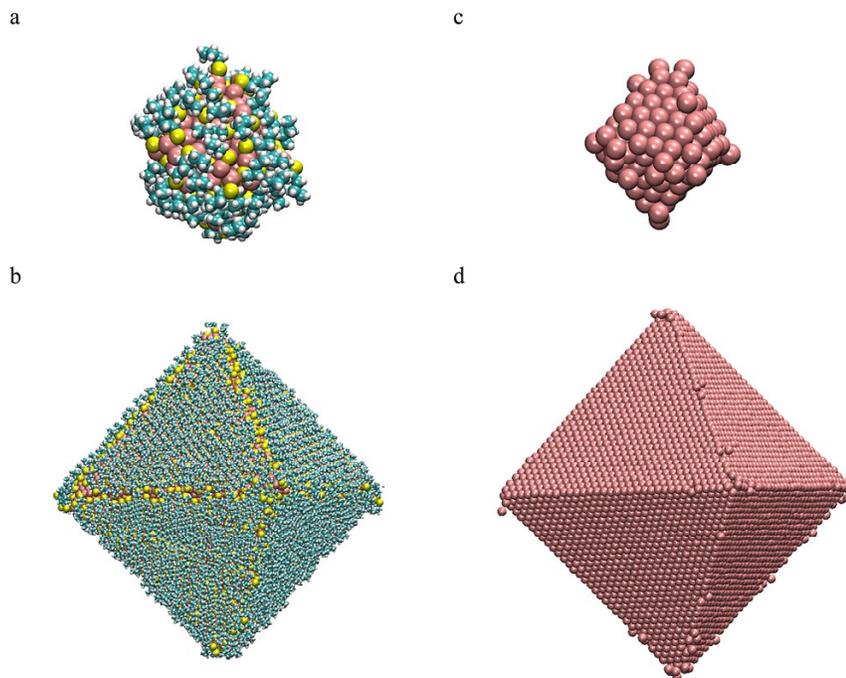

Figure 4: Snapshots of the simulation of butanethiolate-covered silver nanoparticles of 2 nm (a) and 10 nm (b) at the end of the convergence step for 1 ns at 300 K. The ligands are removed to show any reconstruction of the metallic core in (c) and (d) snapshots.

3.2.1. Is the 77 SAM Stable on Nanoparticles?. We now study the frequency at which 77 SAMs are found on the NPs for butanethiolates. For 4 nm, a butane frequency of about 78% is found at low temperatures, as shown in Figure 5 . This is significantly lower than the frequencies between 90 and 100% observed on the flat surfaces (Figure 2 ).

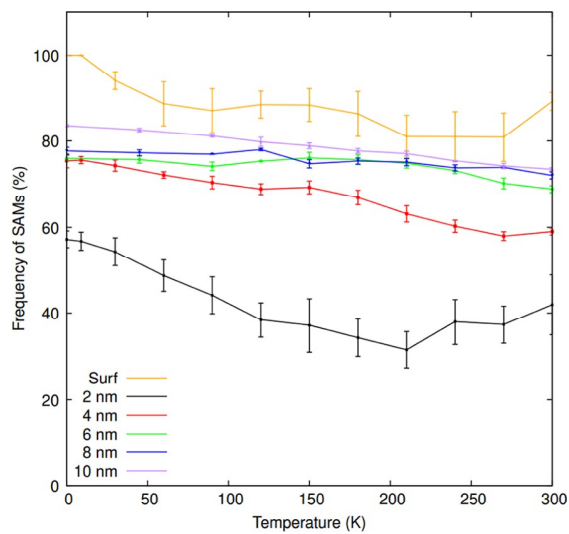

To understand this difference, we studied where the SAMs are located (see Figure 6 ). Therefore, all sulfur atoms belonging to SAMs are colored yellow. They are found principally in the centers of the facets, and the SAM formation is perturbed by the edges.

Figure 5:Frequencies of 77 SAMs on NPs for butanethiolates depend on the temperature. Comparison between surfaces and NPs of different sizes from 2 to 10 nm; equilibration time of 1 ns.

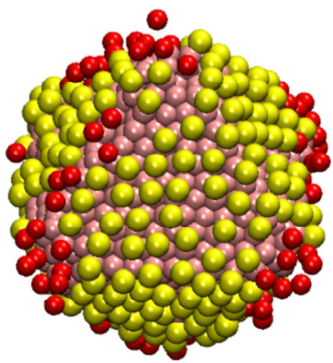

Figure 6:Snapshots of the simulation of 77 butanethiolate SAMs on Ag (111) nanoparticles of 4 nm at the end of the convergence step for 1 ns at 300 K. Atoms belonging to the SAM are colored in yellow

By increasing the temperature, the SAM frequency decreases. A frequency of 77 SAMs of 60% is observed at 300 K.

Let us look at the evolution of the 77 SAM frequency with particle size (see Figure 5). From 2 to 10 nm, the SAM frequency decreases at 300 K for butanethiolate. Nevertheless, depending on the NP diameter, the decrease is more or less pronounced. The more we increase the NP diameter, the more the coverage of the facets grows, and the SAM formation is then less perturbed by the edges.

3.2.2. The Structure of the Metallic Core. We now turn to the number of neighbors observed for the silver atoms. For a perfect octahedron, 7 and 9 neighbors are found on the edges and facets, respectively. A change in these numbers would indicate a metallic restructuring due to the ligands, for example, on the edges. This local restructuring should be distinguished from the global one observed on silver surfaces in experiments. As explained in Section 3. Such general restructuring is impossible to see during ReaxFF simulations. Such a phenomenon was widely observed in the ReaxFF simulation for gold NPs, where the metallic atoms were extracted on the facets and, in particular, at the edges due to staple formation. 30,31 Figure 7 shows the frequencies of silver atoms as a function of their number of neighbors for different NP sizes with butanethiolate SAMs. In the Supporting Information (Figure S10), the neighbor frequencies of 4 nm NPs are plotted for butanethiolate at different simulation times. The frequencies of neighbors agree for the three runs of different lengths, showing that these results are well converged.

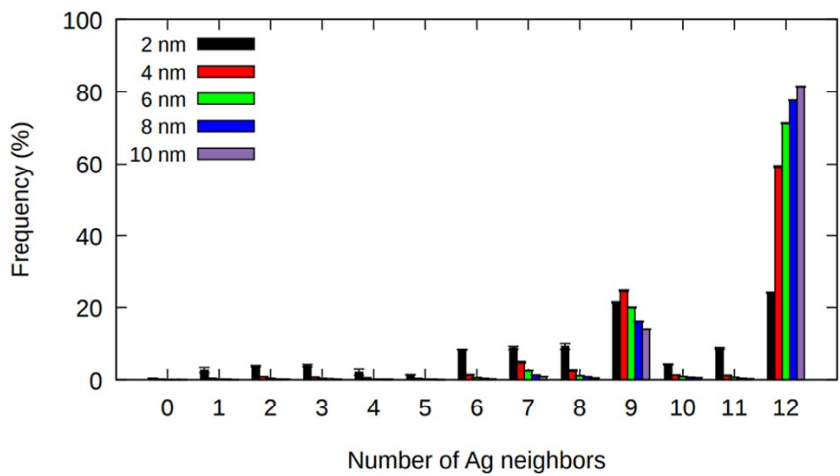

Figure 7: Frequencies of silver atoms as a function of the number of neighbors for NPs from 2 to 10 nm for butanethiolate (equilibration time of 1 ns and convergence time of 1 ns). The property was taken at the end of the convergence at 300 K.

By focusing on Figure 7, it can be seen that the neighbor frequencies for the NPs larger than 2 nm are similar to those of a perfect octahedron, with peaks for 7, 9, and 12 neighbors. Only at the edges do some atoms change their positions leading to atoms with 6 or 8 neighbors. These results are very different from the ones found for gold NPs previously discussed.31 This shows that no surface restructuring spontaneously happens for large silver NPs.

However, for the 2 nm NP, the behavior is markedly different (see black lines in Figure 6). A large fraction of atoms changes the number of neighbors, even on the facets. This change in morphology is confirmed in the snapshots shown in the Supporting Information (Figure S8).

3.2.3. Location and Occupation of Adsorption Sites. We first examine where the ligands are located on the NP surface. Ligands are counted for the vertices or edges if they are in contact with at least one silver atom on the vertices or edges, respectively. The remaining ligands are attributed to the facets. For example, in the case of 4 nm particles with butanethiolate, 55.7, 39.4, and 4.9% of the ligands are adsorbed on the facets, edges, and vertices, respectively. These frequencies do not significantly change during the simulation and correspond to the frequencies initially set up. Similar behavior was observed for all NP sizes and ligands. In Table S2, all frequencies giving the locations of ligands are collected.

We now turn to the question of which adsorption sites the ligands occupy. In Figure 8, the frequencies of head groups as a function of their number of silver neighbors are shown for butanethiolate and different NP sizes. On a perfect Ag (111) surface, the headgroup of a ligand occupying the on−top, bridge, or hollow site has 1, 2, or 3 neighbors, respectively. For NPs larger than 4 nm, this is also true here since the degree of silver atoms extracted by ligands is very low, as previously shown, and the NP surface is well described by the Ag (111) plane. The frequencies of neighbors agree for the three runs with different simulation lengths (see Figure S11). Similar to the surface results discussed in Section 3.1, the ligands on the NP facets occupy only on−top and bridge sites, with a preference or the on−top site. Also, the NP size has little influence on the occupation frequencies, except for the smallest NP of 2 nm. On the NP's edges, the results are quite different where the head groups are mainly in contact with two atoms, which indicates a preference or the bridge site.

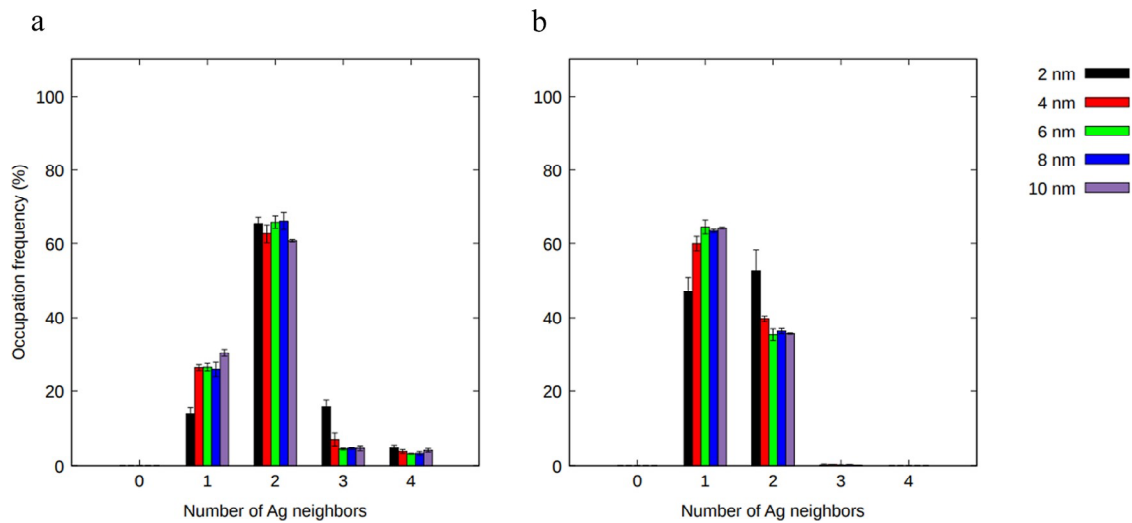

Figure 8: Occupation frequencies of S atoms in contact with a given number of silver atoms on the (a) edges and (b) facets for butanethiolates on the Ag(111) nanostructure from 2 to 10 nm with a simulation length of 1 ns at the end of the convergence at 300 K.

The occupation of the on−top sites found on the facets and edges is in good agreement with recent S−edge XANES experiments,17 which clearly show an occupation of this site. Here, we show that this occupation is related to a dense assembly such as the 77 SAM, which can be realized only by occupying both on−top and bridge sites. This explains the occupation of the on−top sites even when the bridge sites are preferred, as shown in DFT calculations. 18,19 In contrast, less dense assemblies such as the 33 SAM would imply only an occupation of the bridge site that is not consistent with experiments.

In the gold NP ReaxFF simulations, an on−top site preference was also observed on the facets. In good agreement with experiments for gold NPs.31 A large fraction of ligands was also in contact with three or even four gold atoms, which is not the case for the simulation obtained here for silver NPs. Please note that, in particular, for the synthesis of NPs, first thiols are physisorbed, which leads to the formation of thiolate with a strong bond between the ligand and the silver atom. What happened with the H atoms of the thiols is still an open question; it may be adsorbed by the metal. In our work, the H atoms are ignored since they are much smaller than the thiolates and may easily fit in the voids between the ligand head groups. We have carried out simulations for the 77 SAM placed on the silver surface H atoms between sulfur atoms. The frequency of SAMs is reduced by about 20%, which shows that these hydrogens may perturb the assembly.

Finally, we study the average distances between neighboring sulfur head groups of the ligands. The S–S distances are shown in Figure 9 for butanethiolate and different NP sizes. The S–S distance is close to 4.4 Å for both the edge and the facet, as expected for a 77 SAM. For NPs larger than 4 nm, this distance does not evolve with the NP size or alkane chain length of the ligand. There is also no significant difference between the edge and the facet. In a previous simulation of gold NPs using nonreactive force fields, 39,40 the S–S distances on the edges were found to be about 0.2 Å smaller than those on the facets, which implies a different assembly on the edges. These results could not be confirmed by recent ReaxFF simulations of gold NPs. Here, we show that, also for silver, the S–S distance on the edges is not reduced. For smaller NPs of 2 nm, the average S–S distance increases to 4.5 nm, and the distances on the edges are slightly reduced with respect to the facets.

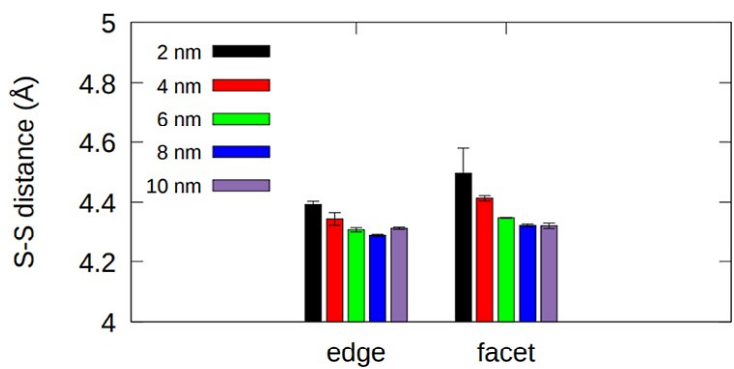

Figure 9: Average S–S distance between first neighbors on the edges and facets of silver NPs from 2 to 10 nm for butanethiolates at the end of convergence at 300 K for an equilibration length of 1 ns.

## 4. Conclusions

We have shown here that the ReaxFF approach is able to yield a stable 77 SAM typical of butanethiolates on the Ag (111) surface, demonstrating its effectiveness in simulating and predicting surface adsorption behaviors with high accuracy. It is interesting to note that ReaxFF used in a recent paper to study gold NPs has shown completely different behaviors with the formation of staples on the Au (111) surfaces. The formation of staples is in good agreement with the experiment and the DFT calculation.19 This shows that ReaxFF yields different properties for both materials in contrast to other force fields. In the recent paper, it was also shown that ReaxFF was capable of reproducing the formation of staples on Au (111) surfaces. Thus, it is possible with ReaxFF to distinguish both materials, which is usually quite difficult for FFs since both metals are characterized by the same geometry. However, we observed that the 77 SAM of methanethiolate is not stable for very long simulations. In addition, the SAM for butanethiolate is not energetically favored with respect to other SAMs. This shows the model's current parametrization limits. Moreover, the reconstructed surface usually observed for silver is made of silver adatoms on an Ag (111) surface. This corresponds to an important restructuring with respect to our systems, which is impossible to see during the ReaxFF simulations. But our simulations do not even show any perturbation of the silver layer in contact with the thiolates. This indicates the good stability of these unrestructured SAMs. It would be, however, interesting in the future to study already restructured silver surfaces with thiolates; the idea would be to compare the binding energies of these systems with the unrestructured ones obtained here. This would allow us to see which one is energetically preferred and if restructuration can be experimentally expected.

The simulations here also show that dense assemblies such as the 77 SAMs found on the Ag (111) surface should appear on silver NPs. To the best of our knowledge, there are no available experiments studying this question. During the simulations, no decomposition of the alkanethiolate was found. The origins of the silver sulfide observed in some experiments are an open question. 11,12,17 In the future, this question should also be investigated for restructured surfaces and other ligands.

Also, no spontaneous restructuring of the NP surfaces was found, except for the small NPs inferior to 4 nm. A global restructuring of the silver surfaces, as observed in experiments, is not attainable with ReaxFF simulations. In the future, we plan to perform simulations starting from a restructured surface and compare its binding energies with those found here.

Two adsorption sites on silver NP facets were found: one on−top and another at the bridge site, where the former is preferred: this is in good agreement with recent experiments.

Regarding the distance between S−headgroup ligands, we observed no difference between the facets and the edges.

Acknowledgments. This work was granted access to the HPC resources of CINES/IDRIS/TGCC under the allocation 2021-A0080811426 (Responsable: J. Richardi) made by GENCI.

Supporting Information

Section to discuss details of the paper: Ag(111) layer with the initial positions of the sulfur atoms in a 77 SAMs of thiolates. FILE "SItxtartncsilver" (PDF)

Average distances between neighboring S head groups for silver nanocrystals for both methane and butane thiolates, relative frequencies of the S headgroup as a function of their adsorption position for silver nanocrystals for both methane and butane thiolate, frequencies of silver atoms for silver nanocrystals for both methane and butane thiolate, and frequencies of S atoms in contact with a given number of silver atoms for silver nanocrystals for both methane and butane thiolate. FILE "SItxtartncsilver" (PDF)

Frequencies of 77 and 33 SAMs of methane and butane thiolates on Ag(111) surface, frequencies of 77 SAMs of methane and butanethiolates on Ag(111) surface after equilibration, snapshot of both the methane and butanethiolates SAMs on Ag(111) surface after 4 ns, difference of binding energies between the 77 and the 33 SAMs for both butane and methanethiolates, frequencies of atoms as a function of the number of Ag neighbors for butanethiolates at different simulation lengths from 0.3 to 3 ns, occupation frequencies of adsorption sites on the facet for butanethiolates on Ag(111) surface, average distance between sulfur atoms neighbors of butanethiolates on a 77 SAM, snapshots of the simulation of methanethiolate covered silver nanoparticles of 2 and 4 nm at the end of the convergence step for 1 ns at 300 K, frequencies of 77 SAM for both methane and butanethiolates SAMs on a 4 nm NP, frequencies of silver atoms as a function of their number of neighbors for 4 nm for butanethiolate at different equilibration times, and occupation frequencies of adsorption sites on the edges and facets for butanethiolates on a 4 nm NP. FILE

"SIfigartncsilver"(PDF)

Program for setup of surface conformation; FILE "SAM_maker.py" (TXT)

Program for setup of nanocrystals conformation; FILE "ntm_setup.py" (TXT)

Methane_Thiolate_RightShift.xyz (XYZ)

Program for analysis of nanocrystal and surface simulations; FILE "ntm_ana.py" (TXT)

Methane_Thiolates_straight.xyz (XYZ)

Butane_Thiolate.xyz (XYZ)

nc6octa.xyz (XYZ)

Methane_Thiolate.xyz (XYZ)

simu_ns6_SMe.xyz (XYZ)

User guide for the programs; FILE "readme.txt" (TXT)

ns6Ag287_77N.xyz (XYZ)

The authors declare no competing financial interest.

# ReaxFF Simulations of Self-Assembled Monolayers on Silver Surfaces and Nanocrystals

A. Lahouari,[1] J.−P. Piquemal,[1] and J. Richardi[1]*

*1 Sorbonne Université, Laboratoire de Chimie Théorique, UMR 7616 CNRS , Paris 75005 , France*

*Email johannes.richardi@sorbonne−universite.fr

Supporting Information, Figures

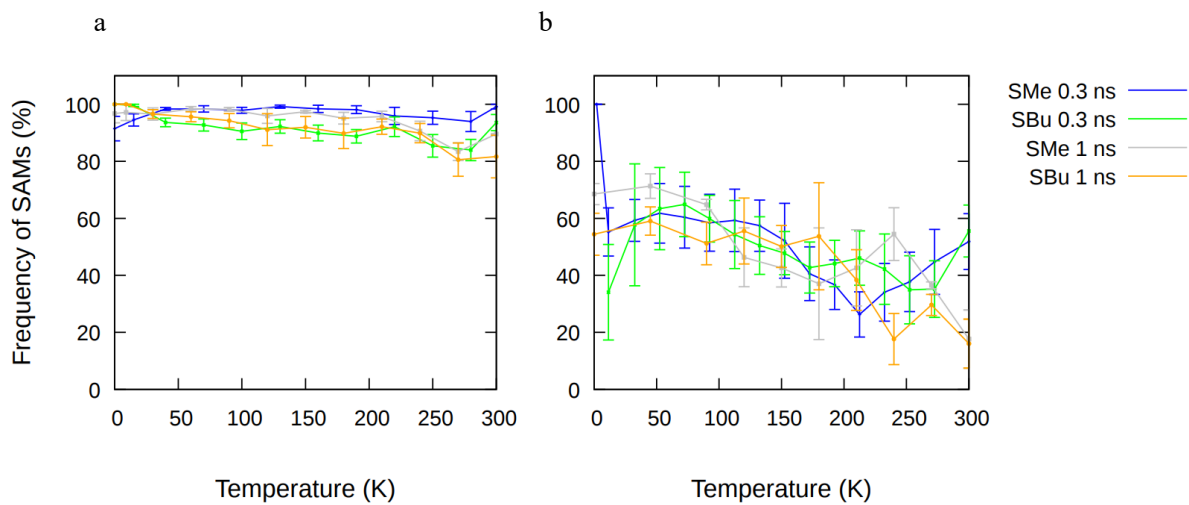

*Figure S1: Frequencies of 77 (a) and 33 (b) SAMs of methane and butane thiolates on Ag(111) surface for an equilibration of 0.3 and 1 ns.*

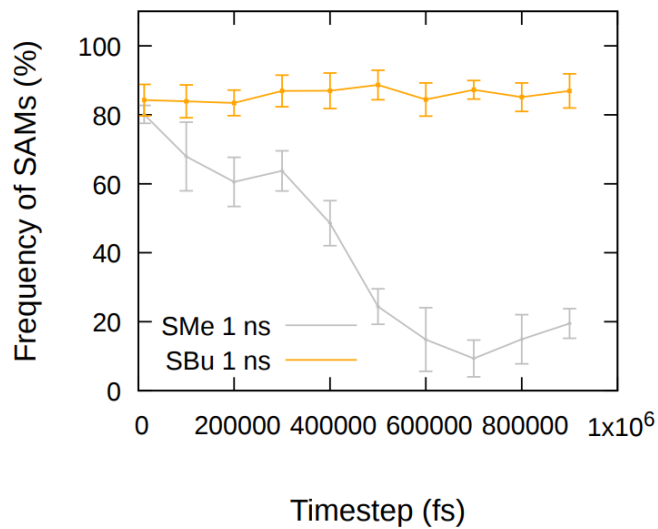

**Figure** *S2: Frequencies of 77 SAMs of methane and butane thiolates on Ag (111) surface continuing the run after the equilibration of 3 ns for 1 ns.*

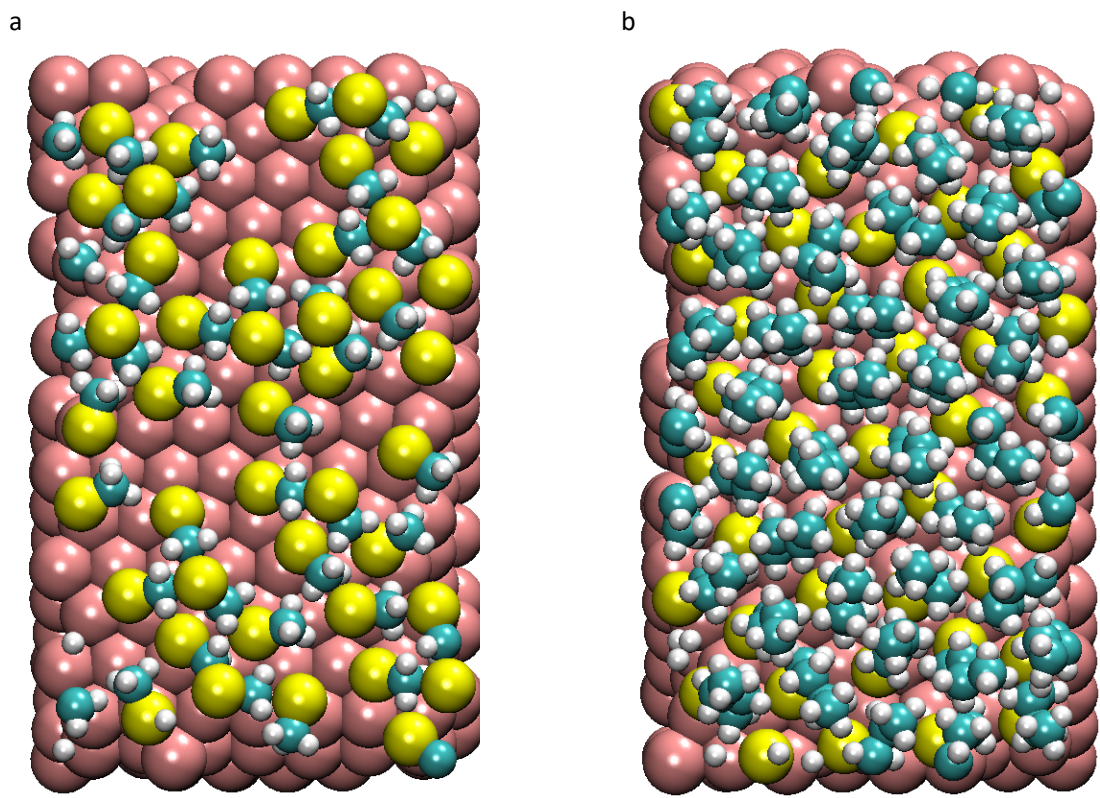

*Figure S3: Snapshot of both the methane (a) and butane (b) thiolates SAMs on Ag(111) surface after 4 ns.*

a b

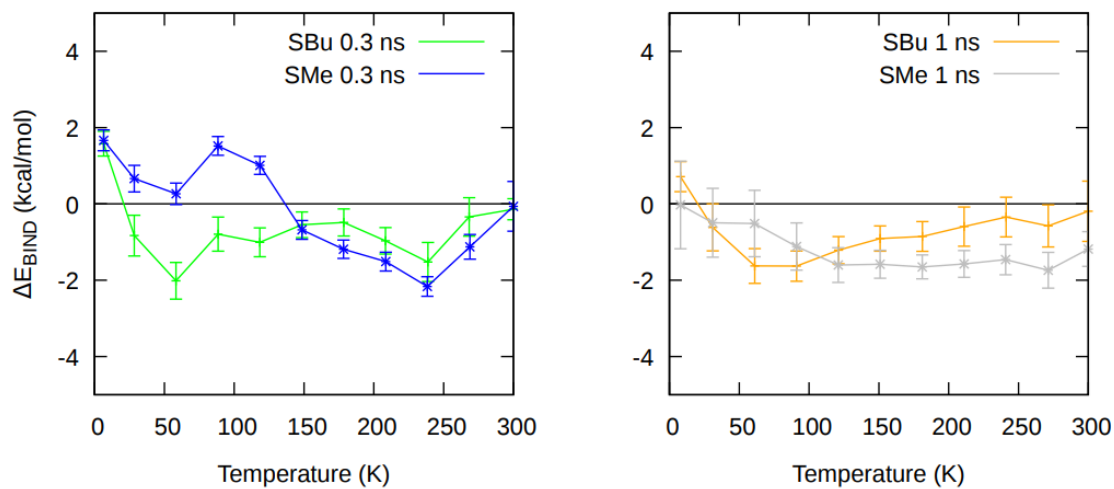

*Figure S4: Difference of binding energies between the 77 and the 33 SAMs as a function of the temperature for both butane and methane thiolates on the Ag (111) surface. Simulation length: (a) 0.3 ns and (b) 1 ns.*

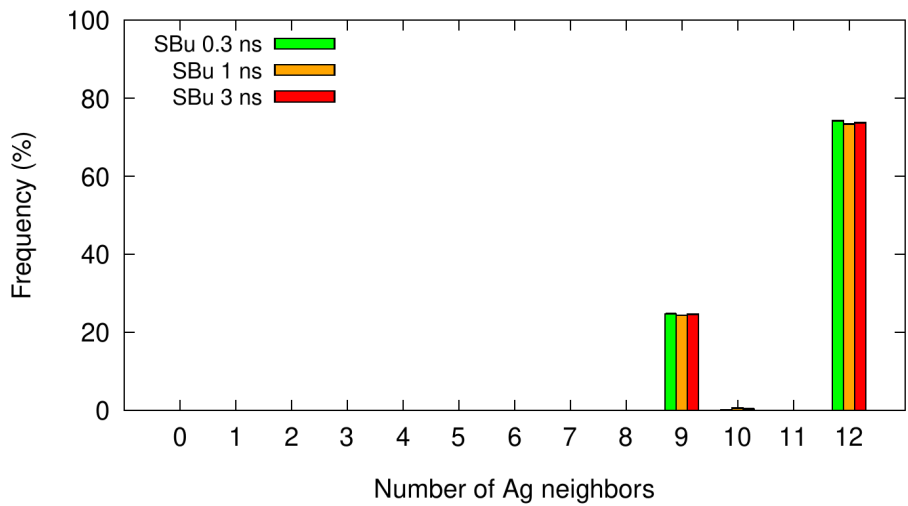

*Figure S5: Frequencies of atoms as a function of the number of Ag neighbors for an Ag (111) surface with a SAM composed of butane thiolates at different simulation length from 0.3 ns to 3 ns.*

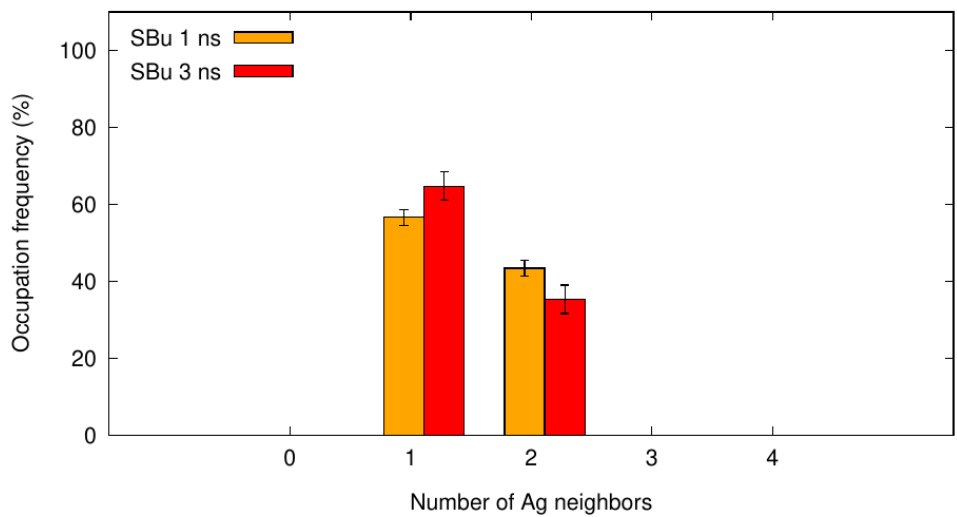

*Figure S6: Occupation frequencies of adsorption sites on the facet for butane thiolates on Ag (111) surface at different simulation lengths, where respectively 1, 2 and 3 Ag neighbors correspond to on-top, bridge and hollow sites.*

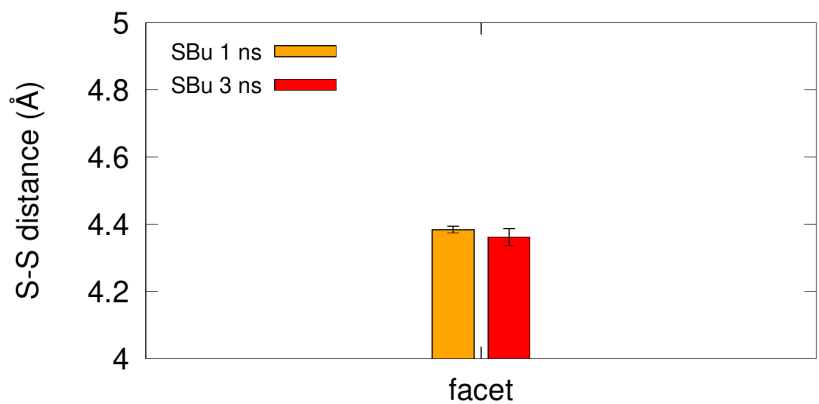

*Figure S7: Average distance between sulfur atoms neighbors of butane thiolates on a 77 SAM at different simulation lengths 1 ns and 3 ns. Result at the end of the equilibration at 300 K.*

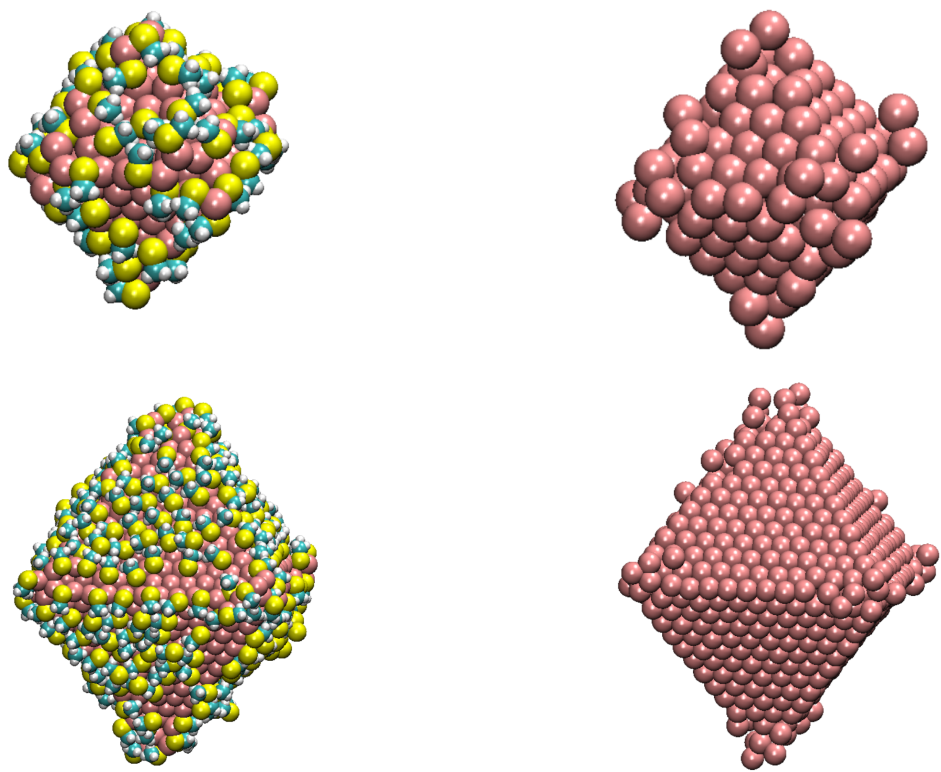

*Figure S8: Snapshots of the simulation of methane thiolate covered silver nanoparticles of 2 nm (a) and 4 nm (c) at the end of the convergence step for 1 ns at 300 K. The ligands are removed to show any reconstruction of the metallic core on b and d snapshots.*

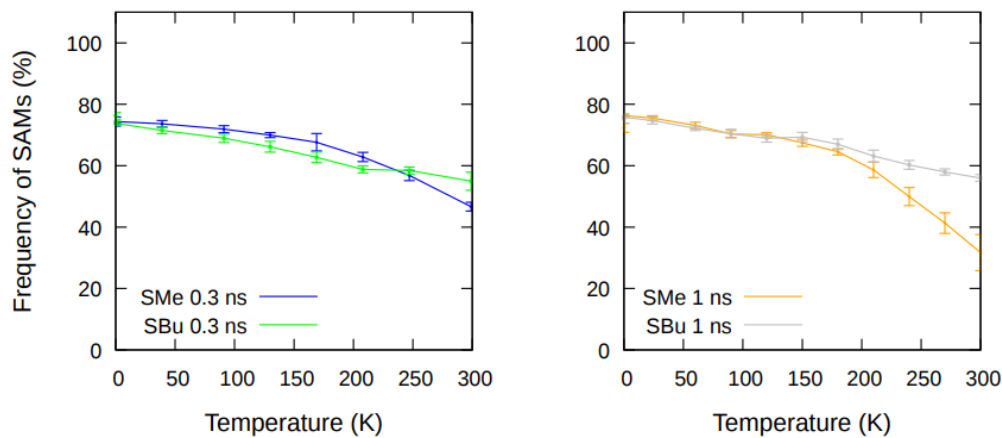

*Figure S9: Frequencies of 77 SAM for both methane and butane thiolates SAMs on a 4 nm Ag (111) NP for different for 0.3 ns, 1 ns and 3 ns simulation length during the equilibration step.*

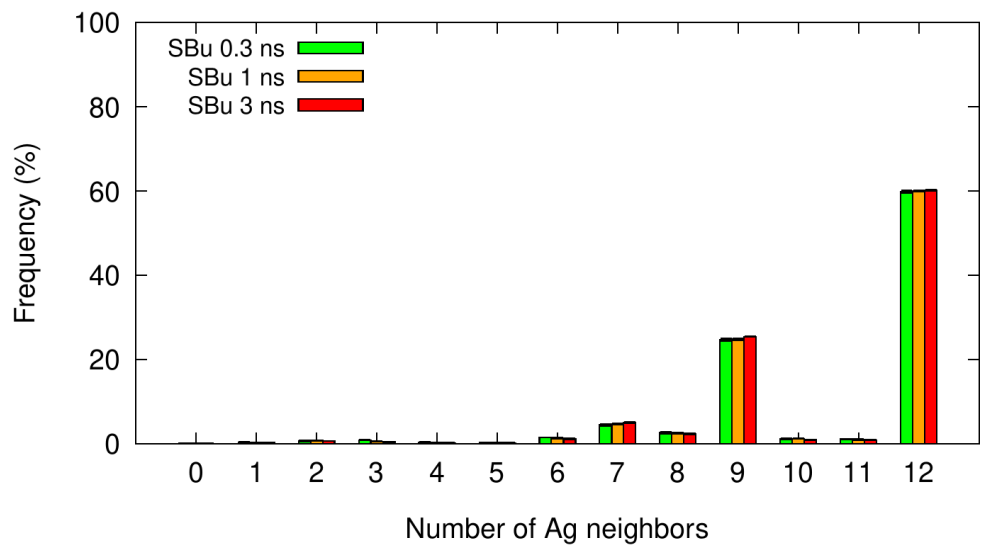

*Figure S10: Frequencies of silver atoms as a function of their number of neighbours for 4 nm for butane thiolate at different equilibration times from 0.3 ns to 3 ns. Property taken at the end of the convergence at 300K.*

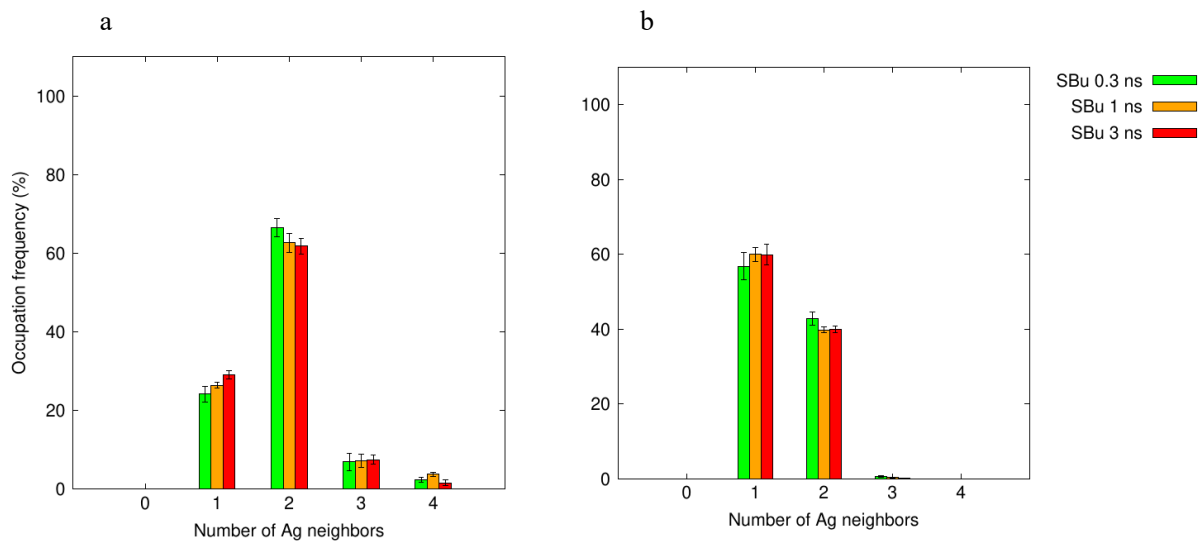

*Figure S11: Occupation frequencies of adsorption sites on the (a) edges and (b) facets for butane thiolates on an Ag (111) nanoparticle of 4 nm, where respectively 1, 2 and 3 Ag neighbours correspond to top, bridge and hollow sites. Results for equilibration times 0.3, 1 and 3 ns at the end of the convergence at 300 K.*

# ReaxFF Simulations of Self-Assembled Monolayers on Silver Surfaces and Nanocrystals

A. Lahouari,[1] J.−P. Piquemal,[1] and J. Richardi[1]*

*1 Sorbonne Université, Laboratoire de Chimie Théorique, UMR 7616 CNRS , Paris 75005 , France*

*Email johannes.richardi@sorbonne−universite.fr

Supporting Information,

1. SAMmaker

The tool SAMmaker generates the positions of the metallic atoms and the ligands for a 33 and 77 SAMs. First, this tool calculates the positions of the metallic atoms of an fcc slab with its highest layers corresponding to the (111) plane. The lines of the metallic atoms building up the (111) plane points into the x direction as shown in the following figure SI1. The user can then define the positions of the ligands for the first lines in the x direction. The headgroups of the ligands for a 77 SAM are colored in yellow in the following figure. These ligands are than repeated in the y direction taking into account a possible shift in the x direction. The user can choose the length of the alkyl chain and also specify the restructured conformation.

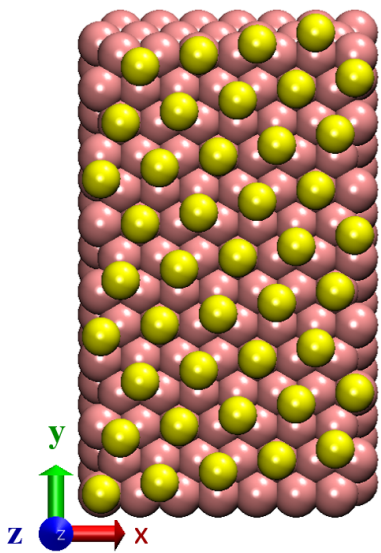

Figure S1: Ag (111) layer with the initial positions of the sulfur atoms in a 77 SAMs of thiolates.

2. SamFinder

In order to determine which ligands, belong to a self-assembled monolayer characterized as 33 or 77 we start as follows. First, a head group atom is picked up by accident. All surrounding head groups at a distance $d_{SAM}$ within a precision of 10% are determined, where $d_{SAM}$ is the distance usually observed for the investigated SAM (e.g. 5.0 Å for 33 and 4.39 Å for 77). Then, the distances between these surrounding atoms are calculated. Whenever one of the surrounding head groups has two neighbors at a distance $d_{SAM}$ within a precision of 10%, it is counted as a new member of the SAM. In the following, new SAM members are chosen as centers and the procedure described above is reiterated. This percolation procedure stops when no newer neighboring atoms belonging to the SAM can be found. It is then restarted with a new head group belonging to no SAM previously found and chosen by accident. This procedure continues until all head groups have been investigated.

3. Entropy

The free energy of a ligand in a SAM related to entropy can be obtained from equation:

$$F = -kT \ln q_t$$

where $q_t$ is the partition function of translation for the ligand with mass m defined as

$$q_t = \frac{V}{h^3} \sqrt{2\pi mkT}$$

V is the volume in which the ligand can move for the SAM.

The difference in free energy between both SAMs can then be calculated from

$$\Delta F = -kT \ln (V_{77}/V_{33})$$

Using the relation $S = -(\partial F/\partial T)$, it can be seen this $\Delta F$ is directly related to the entropy.

It is now assumed that the ligand movement, perpendicular to the metallic surface, does not depend on the type of SAM, but on the metal-ligand interaction. Then the ratio of volumes can be approximated by $V_{77}/V_{33} \simeq S_{77}/S_{33}$ where $S_{77}$ and $S_{33}$ are the surface per ligand for both SAMs. It can be calculated for the 33 and 77 SAMs as 20.37 Å$^2$ and 15.84 Å$^2$. This gives at 300 K for the free energy difference 0.149 kcal/mol which shows that the 33 SAM is entropically favored.

# ReaxFF Simulations of Self-Assembled Monolayers on Silver Surfaces and Nanocrystals

A. Lahouari,[1] J.−P. Piquemal,[1] and J. Richardi[1]*

*1 Sorbonne Université, Laboratoire de Chimie Théorique, UMR 7616 CNRS , Paris 75005 , France*

*Email johannes.richardi@sorbonne−universite.fr

Supporting Informations

**Table S1.** Average distances between neighboring S head groups. The results are given for different silver nanocrystal diameters capped by butane thiolate. The values on the vertex, edges and facets are shown.

```
2 nm / SBu
total    4.44 ± 0.01
vertex   4.39 ± 0.01
edge     4.39 ± 0.01

4 nm / SBu
total    4.43 ± 0.01
vertex   4.37 ± 0.04
edge     4.34 ± 0.02

6 nm / SBu
total    4.35 ± 0.00
vertex   4.34 ± 0.07
edge     4.22 ± 0.01

8 nm / SBu
total    4.31 ± 0.00
vertex   4.32 ± 0.06
edge     4.23 ± 0.00

10 nm / SBu
total    4.29 ± 0.01
vertex   4.41 ± 0.10
edge     4.26 ± 0.01
```

**Table S2.** Relative frequencies of the S head group as a function of their adsorption position for silver nanocrystals with butane thiolate.

```
2 nm / SBu
 vertex          edge           facet
16.99 ± 0.63   65.77 ± 1.06   17.26 ± 0.88

4 nm / SBu
 vertex          edge           facet
 4.93 ± 0.23   39.37 ± 1.13   55.67 ± 0.96
```

```
6 nm / SBu
 vertex          edge           facet
 2.20 ± 0.06   26.27 ± 0.44   71.57 ± 0.49

8 nm / SBu
  vertex         edge           facet
 1.13 ± 0.09   21.53 ± 0.33   77.33 ± 0.41

10 nm / SBu
  vertex         edge           facet
 0.67 ± 0.07   18.23 ± 0.18   81.07 ± 0.23
```

**Table S3.** Frequencies of silver atoms with a given number of neighbors for silver nanocrystals with butane thiolate. First line denoted by "total" gives the absolute number. All other values are given in %.

```
2 nm / SBu
Neihbors       0              1              2              3              4              5              6
Frequency  0.44 ± 0.11  2.14 ± 0.41   4.59 ± 0.43   3.92 ± 0.20   1.69 ± 0.17   1.78 ± 0.29   6.47 ± 0.63
Neighbors      7              8              9              10             11             12
Frequency  10.51 ± 0.79   8.37 ± 0.54   21.60 ± 0.44   3.70 ± 0.48   8.47 ± 0.35   23.99 ± 0.39

4 nm / SBu
Neihbors       0              1              2              3              4              5              6
Frequency  0.10 ± 0.00  0.30 ± 0.00   0.80 ± 0.10   0.53 ± 0.09   0.37 ± 0.03   0.30 ± 0.06   1.43 ± 0.18
Neighbors      7              8              9              10             11             12
Frequency  4.73 ± 0.13   2.57 ± 0.15   24.87 ± 0.27   1.23 ± 0.09   1.03 ± 0.12   59.97 ± 0.17

6 nm / SBu
Neihbors       0              1              2              3              4              5              6
Frequency  0.00 ± 0.00  0.10 ± 0.00   0.20 ± 0.00   0.13 ± 0.03   0.03 ± 0.03   0.10 ± 0.00   0.40 ± 0.00
Neighbors      7              8              9              10             11             12
Frequency  2.80 ± 0.00   0.73 ± 0.03   20.50 ± 0.06   0.83 ± 0.03   0.50 ± 0.06   71.73 ± 0.09

8 nm / SBu
Neihbors       0              1              2              3              4              5              6
Frequency  0.00 ± 0.00  0.10 ± 0.00   0.17 ± 0.03   0.10 ± 0.00   0.03 ± 0.03   0.03 ± 0.03   0.27 ± 0.03
Neighbors      7              8              9              10             11             12
Frequency  1.50 ± 0.00   0.60 ± 0.00   16.47 ± 0.03   0.63 ± 0.03   0.23 ± 0.03   78.03 ± 0.07

10 nm / SBu
Neihbors       0              1              2              3              4              5              6
Frequency  0.00 ± 0.00  0.00 ± 0.00   0.10 ± 0.00   0.10 ± 0.00   0.03 ± 0.03   0.00 ± 0.00   0.17 ± 0.03
Neighbors      7              8              9              10             11             12
```

```
Frequency  1.03 ± 0.03   0.40 ± 0.00  13.87 ± 0.03   0.50 ± 0.00   0.20 ± 0.00  81.87 ± 0.03
```

**Table S4.** Frequencies of the S atoms in contact with a given number of silver atoms for silver nanocrystals with butane thiolate.

```
2 nm / SBu
Neighbors      0               1                2               3               4
vertex     0.00 ± 0.00   12.51 ± 3.19   70.32 ± 4.78   15.87 ± 2.89    1.30 ± 0.71
edge       0.00 ± 0.00   13.98 ± 1.67   65.36 ± 1.78   15.82 ± 1.78    4.85 ± 0.76
facet      0.00 ± 0.00   47.19 ± 3.65   52.58 ± 5.68    0.23 ± 0.23    0.00 ± 0.00
centre     0.00 ± 0.00    0.00 ± 0.00    0.00 ± 0.00    0.00 ± 0.00    0.00 ± 0.00

4 nm / SBu
Neighbors      0               1                2               3               4
vertex     0.00 ± 0.00   15.71 ± 3.47   76.12 ± 3.06    6.12 ± 0.00    2.04 ± 2.04
edge       0.00 ± 0.00   26.44 ± 0.79   62.62 ± 2.44    7.12 ± 1.70    3.81 ± 0.53
facet      0.00 ± 0.00   59.94 ± 1.94   39.75 ± 0.75    0.31 ± 0.05    0.00 ± 0.00
centre     0.00 ± 0.00    0.00 ± 0.00    0.00 ± 0.00    0.00 ± 0.00    0.00 ± 0.00

6 nm / SBu
Neighbors      0               1                2               3               4
vertex     0.00 ± 0.00    0.00 ± 0.00    0.00 ± 0.00    3.18 ± 1.36   96.82 ± 5.45
edge       0.00 ± 0.00   23.37 ± 0.69   56.69 ± 1.03    6.98 ± 0.57   12.96 ± 1.33
facet      0.00 ± 0.00   64.09 ± 1.91   35.67 ± 1.50    0.24 ± 0.13    0.00 ± 0.00
centre     0.00 ± 0.00    0.00 ± 0.00    0.00 ± 0.00    0.00 ± 0.00    0.00 ± 0.00

8 nm / SBu
Neighbors      0               1                2               3               4
vertex     0.00 ± 0.00    0.00 ± 0.00    0.00 ± 0.00    0.00 ± 0.00  100.00 ± 7.96
edge       0.00 ± 0.00   10.37 ± 0.33   24.19 ± 0.47    3.26 ± 0.93   62.19 ± 1.07
facet      0.00 ± 0.00   47.03 ± 0.66   27.91 ± 0.84    0.99 ± 0.16   24.07 ± 0.37
centre     0.00 ± 0.00    0.00 ± 0.00    0.00 ± 0.00    0.00 ± 0.00    0.00 ± 0.00

10 nm / SBu
Neighbors      0               1                2               3               4
vertex     0.00 ± 0.00    0.00 ± 0.00    0.00 ± 0.00    0.00 ± 0.00  100.00 ±10.45
edge       0.00 ± 0.00    2.19 ± 0.33    6.01 ± 0.93    2.73 ± 0.33   89.07 ± 0.33
facet      0.00 ± 0.00   24.49 ± 0.27   15.04 ± 0.18    1.48 ± 0.00   59.00 ± 0.37
centre     0.00 ± 0.00    0.00 ± 0.00    0.00 ± 0.00    0.00 ± 0.00    0.00 ± 0.00
```